# Application of Multivariate Data Analysis to machine power measurements as a means of tool life Predictive Maintenance for reducing product waste

Darren A. Whitaker[1], David Egan[1], Eoin O' Brien[2], David Kinnear[2]

[1]Pharmaceutical Manufacturing Technology Centre, Bernal Institute, University of Limerick, Limerick, Ireland

[2]Department of Design and Manufacturing Technology, School of Science and Engineering, University of Limerick, Ireland

# 1  Abstract

Modern manufacturing industries are increasingly looking to predictive analytics to gain decision making information from process data. This is driven by high levels of competition and a need to reduce operating costs. The presented work takes data in the form of a power measurement recorded during a medical device manufacturing process and uses multivariate data analysis (MVDA) to extract information leading to the proposal of a predictive maintenance scheduling algorithm. The proposed MVDA model was able to predict with 100 % accuracy the condition of a grinding tool.

**Keywords:** Power Measurements; Multivariate Analysis; Predictive Maintenance; Medical Devices; Grinding Tools

# 2  Introduction

Within modern manufacturing and fabrication industries, high levels of competition is a driving force behind an ever increasing desire to reduce overhead costs. Many companies focus on reducing raw material wastage and increasing efficiency. One area in which there is significant potential for cost reduction is machine and tool maintenance management; this has the potential to remove downtime, increase efficiency, reduce defective products and remove equipment inactivity. Overall, this results in a greater quality of product delivered to the consumer. This is of particular importance within the medical device sector where manufacturing costs can be the difference between a life changing product coming to market or not.

Within these industries, machining is a term which describes a broad range of processes which remove material from a workpiece to create the final form. Grinding is a particular case where the characteristics of the cutting tool set it apart from other machining processes. With grinding, thousands of abrasive grains make up the surface of the cutting tool – the grinding wheel. These abrasive grains have irregular shapes and random orientations. In contrast, drilling, milling and turning employ defined cutting edges to remove material from the workpiece using a shearing mechanism. In contrast, the mechanism of material removal in grinding is dependent on the complex interaction between the grinding wheel and the workpiece. As with most cutting tools, over time the grinding wheel becomes dull and the efficiency at which it is able to function is reduced. To ensure that a grinding process operates at maximum efficiency the grinding wheel must be replaced when it is no longer able to carry out its function effectively.

The efficient management of maintenance activities, such as tool replacement, is essential to decrease downtime and defective products(Mobley 2002) in all manufacturing industries. Maintenance management approaches have evolved and can be grouped into three main categories, which in order of increasing complexity and efficiency(Susto et al. 2015), are as follows.

1) Run-to-failure (R2F) or Corrective Maintenance (CM)—the earliest maintenance protocol. Interventions are performed after failures or defects have occurred. This is the most basic approach, while it deals with maintenance it is also the least effective one, as it results in the production of defective parts.

2) Preventative maintenance (PvM)—maintenance actions are carried out according to a planned schedule based on time or process iterations to prevent breakdown. This approach usually results in the prevention of failures, but unnecessary maintenance may also be performed, this is inefficient and can increase the use of resources and operating costs.

3) Predictive Maintainence (PdM)— Maintenance is performed based on an estimate of the " health" status of a piece of equipment(Krishnamurthy et al. 2005). PdM systems enable detection of pending failures and result in timely pre-failure interventions. Prediction tools may be based on historical data, ad hoc defined health factors, statistical inference methods and engineering approaches(Susto et al. 2013).

Predictive maintenance approaches are widely varied, multiple examples exist based on different approaches, such as statistical classification methods(Chao and Tong 2009; Susto et al. 2013, 2015); regression(Hsieh et al. 2013; Susto, Pampuri, et al. 2012); and filtering(Susto, Beghi, et al. 2012; Zhao et al. 2010). Issues of maintenance are different across industries and plants, thus the PdM solution must be tailored for specific problems, explaining the wide variety of approaches in the current literature(Ghahramani 2015).

In this paper we propose a PdM system based on the inferred health of a grinding wheel created from measurements of the power required to perform its function. The assumption being that the tool requires more power to cut as it ages and becomes blunt. This system is based on statistical, data-driven modelling of multivariate data; which in this case is a data stream of power (kW) vs time for the production of a single unit.

## 2.1 Industrial Case Study

Power measurements of a cutting process were taken from a live manufacturing facility where a PvM schedule is in use. This PvM occurs after 2000 process iterations, or 2000 machined workpieces consisting of 5/6 batches, and is used across numerous variations of workpiece dimension. Generally each batch comprises solely of one workpiece dimension. However, batches with different workpiece dimensions, resulting in differing material removal quantities, can be machined using the same tool. As each cutting tool can be used on multiple batches, the 2000 part PvM number is ineffective at describing the true wear on the tool. The varying material removal quantity results in differing levels of wear being subjected to each cutting tool despite them all being used for the same quantity of workpieces. This results in poor cutting tool performance, seen through defective workpieces, which correlates with a R2F maintenance management system under these circumstances.

When the grinding wheel is in a healthy state, no burning is visible on the product. In an unhealthy state burning becomes visible and results in QC failures. It is hoped that a real time analysis of the power trace will result in decision data being provided to aid a more effective maintenance management procedure implemented through a PdM approach.

# 3 Methodology

### 3.1.1 Power Measurements

Power measurements were taken from the cutting tool machine while producing a single product design throughout the complete tool life. To do this, an Emerson USB 2-wire EIA485 converter was used to connect the cutting tool machine to a laptop. The cutting tool machine used in this study was a Rollomatic GrindSmart 620XS. Specialist software included Emerson CTSoft which established a connection between the laptop and machine interface and Emerson CTScope which acted as an oscilloscope, tracing the power of the motor driving the profiled single layer electroplated diamond grinding wheel throughout the grinding cycle. A sampling rate of 50ms was chosen as this gave enough detail to distinguish between the cuts into the work piece. Run data was saved in CSV format for subsequent manipulation in Excel.

### 3.1.2 Experimental

An experiment was undertaken where the power traces of 20 blades were recorded at different times during the complete life of three different grinding wheels, from new to worn enough to cause QC failure such as burning; as detailed in Table 1. For example, with Grinding Wheel 1, power measurements were recorded for 20 units after the wheel had ground 160, 689, 753, 1147 and 1367 units. The burn condition was assessed for each unit.

Table 1 – Cutting wheel specifications

|  | Blades Cut | Burn Condition | Burn rank* |
|---|---|---|---|
| wheel 1 | 160 | none | 1 |
|  | 689 | none | 1 |
|  | 753 | none | 1 |
|  | 1147 | none | 1 |
|  | 1367 | slight, on outer tooth. Started at 1200 parts | 2 |
| wheel 2 | 180 | none | 1 |
|  | 709 | none | 1 |
|  | 774 | none | 1 |
|  | 1125 | none | 1 |
|  | 1400 | slight, on outer tooth. Started at 1300 parts | 2 |
| wheel 3 | 200 | none | 1 |
| | 680 | none | 1 |
| | 900 | none | 1 |
| | 1200 | none | 1 |
| | 1600 | slight, on all teeth, darker on outer. Started at 1260 parts | 2 |

*Burn Ranking: 1 - None, 2 - Small Burn, 3 - Dark Burn

### 3.1.3 Data Analysis

Multivariate data analysis (MVDA) describes the practice of using mathematical and statistical tools to extract information from data tables where each observation contains a large number of variables. In such cases, the desired information lies in the correlation structure between variables, this often leads to erroneous results when tested independently. MVDA by means of projection methods is able to analyse data where challenges such as multidimensionality of the data set, multicollinearity, missing data and variation introduced by deviating factors such as experimental error and noise occur. Principal Component Analysis (PCA) is a commonly used projection method in MVDA, this projects data onto a lower dimensional space where is can be easily inspected. Linear Discriminant Analysis (LDA) is a commonly used technique to classify data into discrete sets *i.e.* yes/no, working/broken or running/faulty.(Nemeth 2003)

*Principal Component Analysis (PCA)*

The most common application of PCA is reducing, with minimal information loss, the dimensionality of datasets(Jolliffe 2002). Typically these data sets consist of large numbers of correlated variables which are transformed into a new set of variables, called Principal Components (PCs). These PCs are uncorrelated and ordered so that most of the variation that was present in the original set of variables is preserved within the first few PCs. For a more detailed description of PCA-based statistical monitoring the interested reader is referred to (Jolliffe 2002) and (Chiang et al. 2000).

*Linear Discriminant Analysis (LDA)*

LDA introduces the concept of supervised classification into data reduction. The aim of LDA is to create a new variable which is a combination of the original predictors by maximizing the differences between the predefined groups, with respect to the new variable. The goal is to combine the predictor scores in such a way that, a single new composite variable, the discriminant score, is formed. This can be viewed as an excessive data dimension reduction technique that compresses the p-dimensional predictors into a one-dimensional line. This linear discriminant function provides a means of discriminating between the groups of data collected during the manufacture of each unit. (Fielding 2007).

Analyses were carried out using the R statistical environment(RDevelopmentCoreTeam 2013) and Rstudio as a graphical front end (RStudioTeam 2015). PCA analysis was carried out using the *prcomp* routine in the core environment and LDA was carried out using the MASS package(Venables and Ripley 2002). Data was imported into the environment using custom routines.

# 4 Results

Typical power curves for the manufacture of a single unit are shown in Fig 1. As the profiled electroplated diamond grinding wheel interacts with the workpiece, the power required by the wheel increases resulting in the power spectrum show. An overlay of the power trace from both a healthy (Black trace) and unhealthy grinding wheel (Red trace) allows a difference to be visualised, however in displaying this data in real time is of little use as it allows no trends to be observed or discerned.

We applied PCA and LDA data analysis to power curves drawn from the industrial process described earlier. It is expected that the power trace will evolve over time, and this evolution can be tracked to predict if the grinding wheel is in a state where no burning is visible (healthy) or in a state where burning is visible (unhealthy).

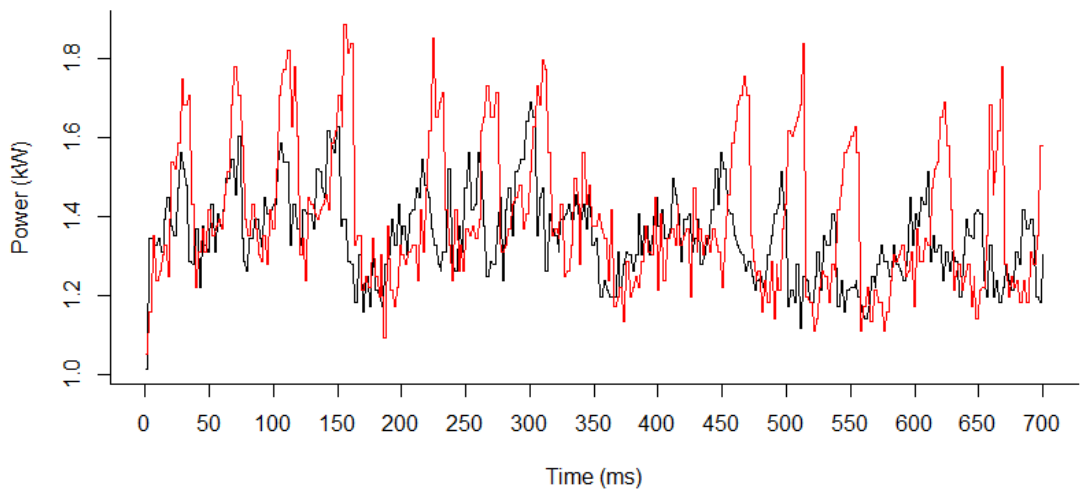

**Figure 1** – Power trace from a healthy (black) and unhealthy (red) grinding disc during a cutting operation

Over the life of a single grinding wheel producing the same product, power measurements for about 20 consecutive units were taken at various stages. By reducing the dimensionality of the power trace, a single descriptor variable, in this case the second principal component (PC2), can be plotted as a function of observation number – which is the power trace taken for each measurement. (Figure 2). This clearly shows that the power reading provides an indicator of evolution over time. Additionally the three right-most points in figure 2 are quite clearly separated from the rest. These points had been identified as the beginning of burning on the product whereas prior to this point burning was not observed. As PCA is unsupervised it is beyond the scope of the technique to report this separation.

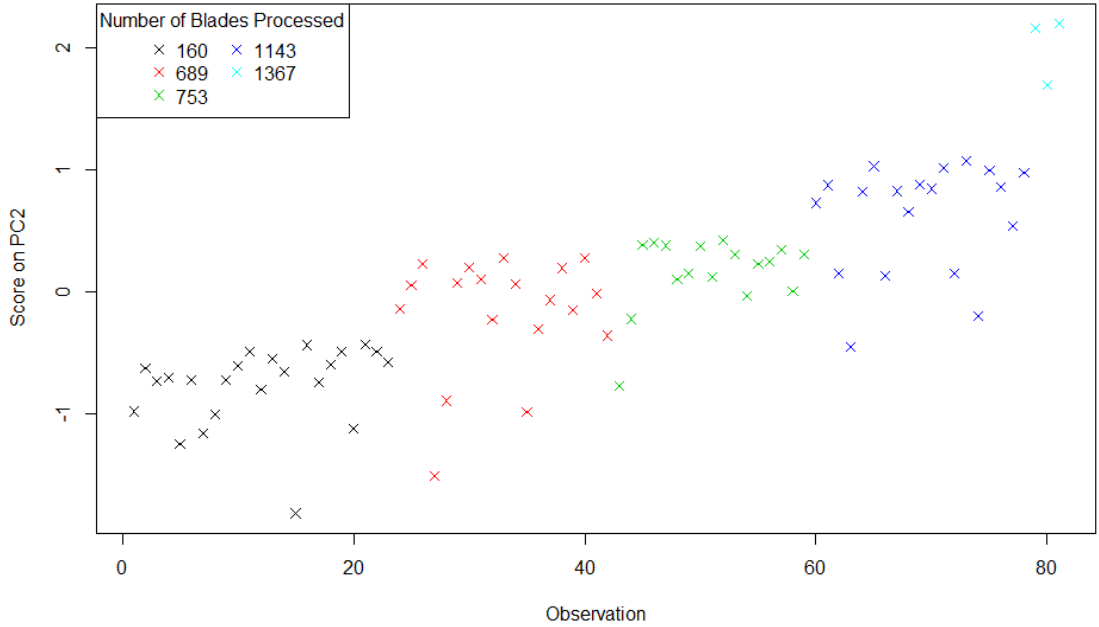

**Figure 2** – PCA analysis showing evolution of PC2 over time

In order to gain some qualitative information from the measurements, a second multivariate technique – linear discriminant analysis (LDA), was employed. Using data from the lifetime of one grinding wheel (Wheel 1) along with knowledge of some " wear criteria" (i.e. No-Burn or Burn) at each point a simple model which returns the criteria can be built. Plotting the first descriptive component, LD1, against observation number (Figure 3) shows that, as in PCA, the trend can be observed and a decision boundary is able to separate the classes.

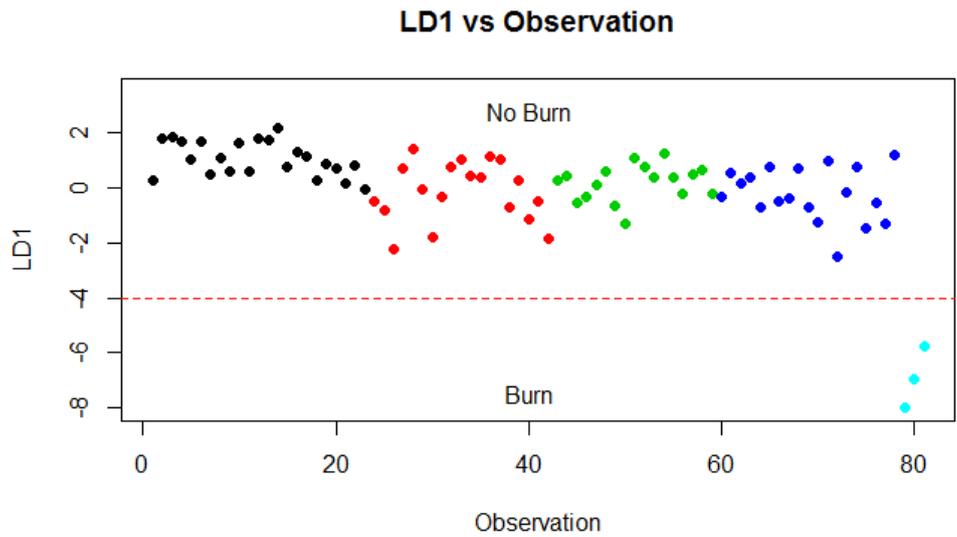

**Figure 3** – LDA analysis showing class separation of observations. Dashed line represents decision boundary. Colours as previous

Predictions of the wear criteria, burn or no-burn, of the power measurements for the remaining grinding wheels (Wheels 2 & 3) can be made by evaluating them using the LDA model. (Table 1). Each observation (i.e. the power spectrum for the grinding of a single unit) is input into the model which then uses the LD1 to classify that observation as either "No-Burn" or "Burn". In each case the model correctly classifies each new observation.

Table 2 – Predicted class of observations (actual class in **bold**, prediction in *italic*) of wheels 2 and 3

|  | Wheel 2 | |
| --- | --- | --- |
|  | **No Burn** | **Burn** |
| *No Burn* | 66 | 0 |
| *Burn* | 0 | 3 |

|  | Wheel 3 | |
| --- | --- | --- |
| *No Burn* | 47 | 0 |
| *Burn* | 0 | 3 |

The predictive system as defined, is an offline model based on historical measurements. In order to allow useful information to be gained, that LDA model would be deployed on a computer connected to the machine to evaluate data in real time (Figure 4). This would allow the power data spectrum for each tool being ground to be input into the model and converted into a tool-health class defining the state of the grinding wheel.

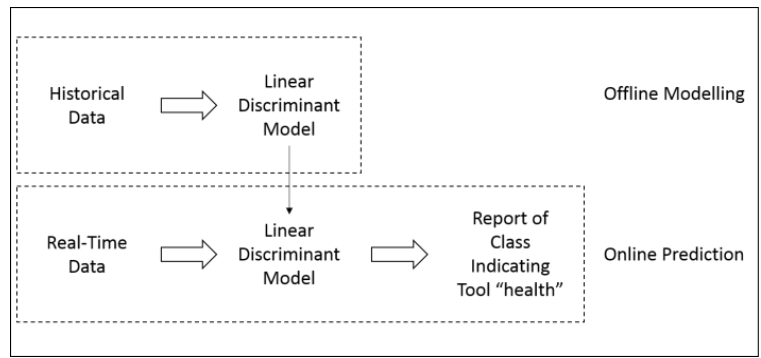

Figure 4 – Approach for simple classification model

Further benefit of this approach is gained by setting control limits which use the value of the linear discriminant function to predict that failure is close (Figure 5). Similar to above, the power spectrum is measured in real-time and input into the model. The LDA output (LD1) firstly results in a tool-health class, but the LD score value can also be used to offer a prediction of whether the tool is nearing its end of life, before burning is actually observed. As there is a defined evolution in the value this would be a simple yet effective approach of alerting operators that the tool is approaching end of life.

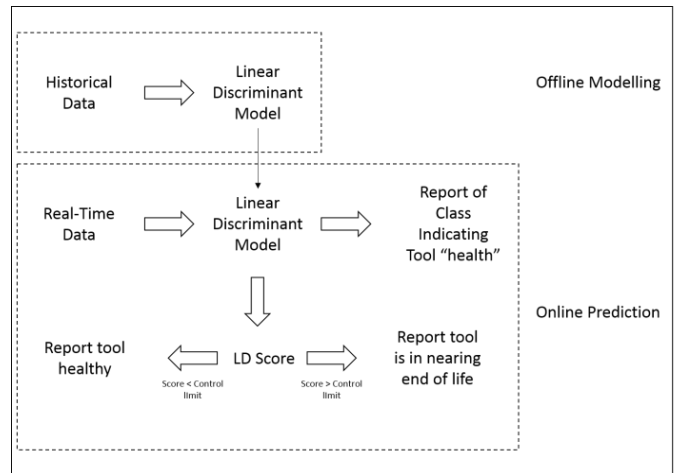

**Figure 5** – Approach for model with predictive capacity

# 5 Conclusions

We present a computationally simple approach to a PdM module for a grinding process which could be applied to almost any process where tools age and the effect resulting in a defective product. By using simple algorithms (PCA, LDA) the module is able to be utilised on the majority of hardware available to plants today.

Further work may look at modelling the typical wear properties of a cutting tool that a more effective maintenance management procedure may be implemented through a more advanced, stochastic PdM approach.

# 6 Acknowledgements

This work was co-funded by the Pharmaceutical Manufacturing Technology Centre (PMTC) under Enterprise Irelands (EI) - Technology Centres Programme & by the European Regional Development Fund (ERDF) under Ireland's European Structural and Investment Funds Programmes 2014 - 2020